\documentclass{PoS}
\pdfoutput=1
\newcommand{\be}{\begin{eqnarray}}
\newcommand{\ee}{\end{eqnarray}}

\title{Positronium: an illustration of nonperturbative renormalization in a basis light-front approach}

\ShortTitle{Positronium: an illustration of nonperturbative renormalization in a basis light-front approach}

\author{\speaker{Xingbo Zhao}$^{1,2,a}$, Kaiyu Fu$^{1,2,b}$, Hengfei Zhao$^{1,2,c}$, James P. Vary$^{3,d}$\\
        $^1$Institute of Modern Physics, Chinese Academy of Sciences, Lanzhou 730000, China\\
        $^2$School of Nuclear Science and Technology, University of Chinese Academy of Sciences, \\Beijing 100049, China\\
        $^3$Department of Physics and Astronomy, Iowa State University, Ames, IA 50011, USA\\
        E-mail: $^a$\email{xbzhao@impcas.ac.cn}, $^b$\email{kaiyufu@impcas.ac.cn}, $^c$\email{zhaohengfei@impcas.ac.cn},
        $^d$\email{jvary@iastate.edu}
}

\abstract{We calculate the mass spectrum and the structure of the positronium system at a strong coupling in a basis light-front approach. We start from the light-front QED Hamiltonian and retain one dynamical photon in our basis. We perform the fermion mass renormalization associated with the nonperturbative fermion self-energy correction. We present the resulting mass spectrum and wave functions for the selected low-lying states. Next, we apply this approach to QCD and calculate the heavy meson system with one dynamical gluon retained. We illustrate the obtained mass spectrum and wave functions for the selected low-lying states.
}

\FullConference{Light Cone 2019 - QCD on the light cone: from hadrons to heavy ions - LC2019\\
		16-20 September 2019\\
		Ecole Polytechnique, Palaiseau, France}

\begin{document}

\section{Introduction}

Basis Light-front Quantization (BLFQ) has been developed as a nonperturbative approach to relativistic bound states~\cite{Vary:2009gt}. It is based on the Hamiltonian formalism and the light-front quantum field theory. In BLFQ, the bound state problem is cast into an eigenvalue problem of the Hamiltonian:
\begin{equation}
  P^{-}|\beta\rangle = P^{-}_{\beta}|\beta \rangle,
\end{equation}
where the eigenvalues $P^{-}_{\beta}$ correspond to the mass spectrum and the eigenvectors $|\beta \rangle$ encode their structural information. In this paper, we report our recent progress in applying BLFQ to the positronium system in QED and the heavy meson systems in QCD.

\section{Positronium}
The positronium (``Ps'') is arguably the simplest bound state system in QED. In this work, we solve the positronium system from first principles - the QED Lagrangian~\cite{Fu:2020b}. In order to make the numerical calculation feasible, we perform basis truncation by retaining the two leading Fock sectors, that is, $|\rm{Ps}\rangle= a|e^+ e^-\rangle+ b|e^+ e^- \gamma\rangle$. In addition, we truncate the basis in the transverse (longitudinal) direction with the truncation parameter $N_{\rm{max}}$ ($K$)~\cite{Vary:2009gt}. Larger $N_{\rm{max}}$ ($K$) translates to more complete bases in the transverse (longitudinal) direction.
We obtain our light-front QED Hamiltonian from the QED Lagrangian via the Legendre transformation. In our truncated basis the light-front QED Hamiltonian, using light-front gauge, takes the following form,
\begin{equation}
P^-_{\rm{QED}}= \int \mathrm{d}^{2} x^{\perp} \mathrm{d} x^{-} \frac{1}{2} \bar{\Psi} \gamma^{+} \frac{m_{e0}^{2}+\left(i \partial^{\perp}\right)^{2}}{i \partial^{+}} \Psi+\frac{1}{2} A^{j}\left(i \partial^{\perp}\right)^{2} A^{j} +e j^{\mu} A_{\mu}+\frac{e^{2}}{2} j^{+} \frac{1}{\left(i \partial^{+}\right)^{2}} j^{+} ,
\label{QEDHami}
\end{equation}
where $\psi$ and $A_\mu$ are the fermion and gauge boson field operators, respectively, and $j^\mu=\bar{\Psi}\gamma^\mu \Psi$.
The first two terms are their corresponding kinetic terms and the remaining terms describe their interaction. $m_{e0}$ is the bare fermion mass. For numerical convenience, we take an artificially increased electromagnetic coupling constant $\alpha=0.15$.

\begin{figure}
\centering
\includegraphics[width=0.45\columnwidth
]{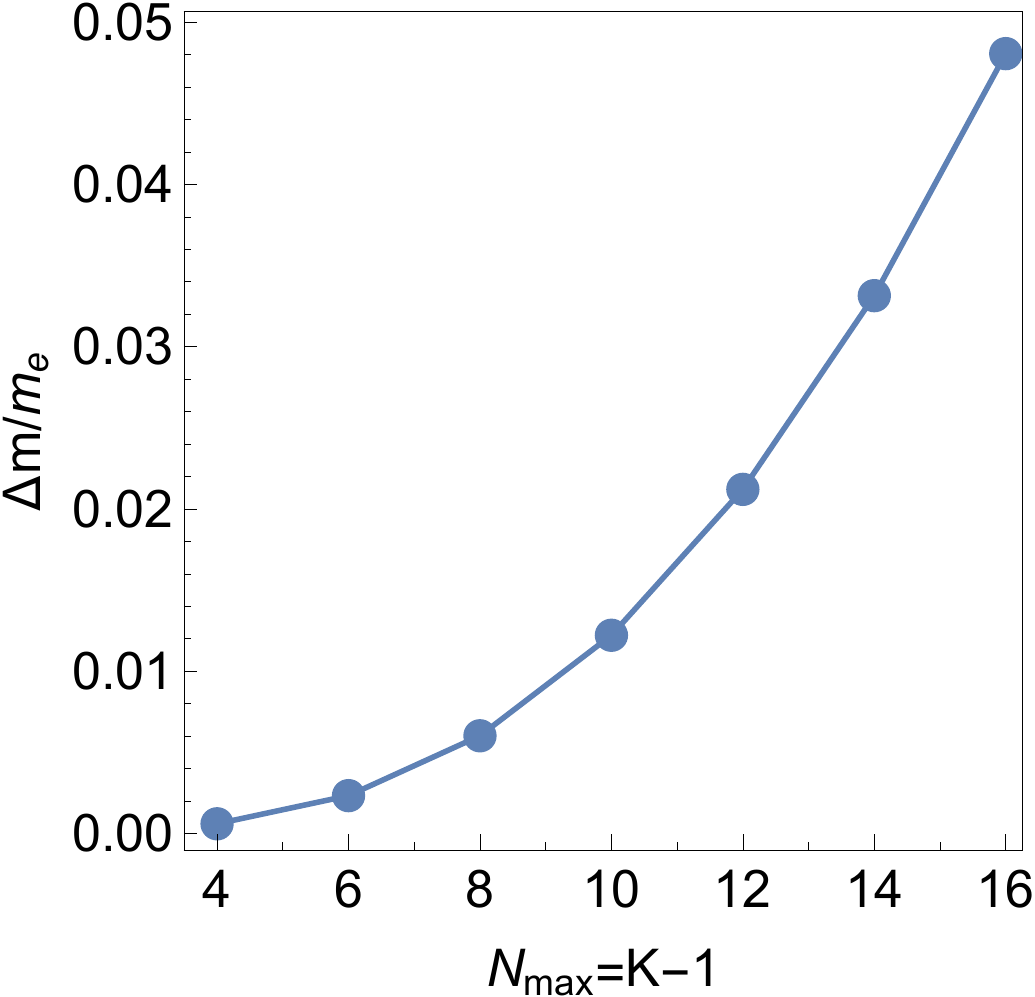}
\includegraphics[width=0.45\columnwidth
]{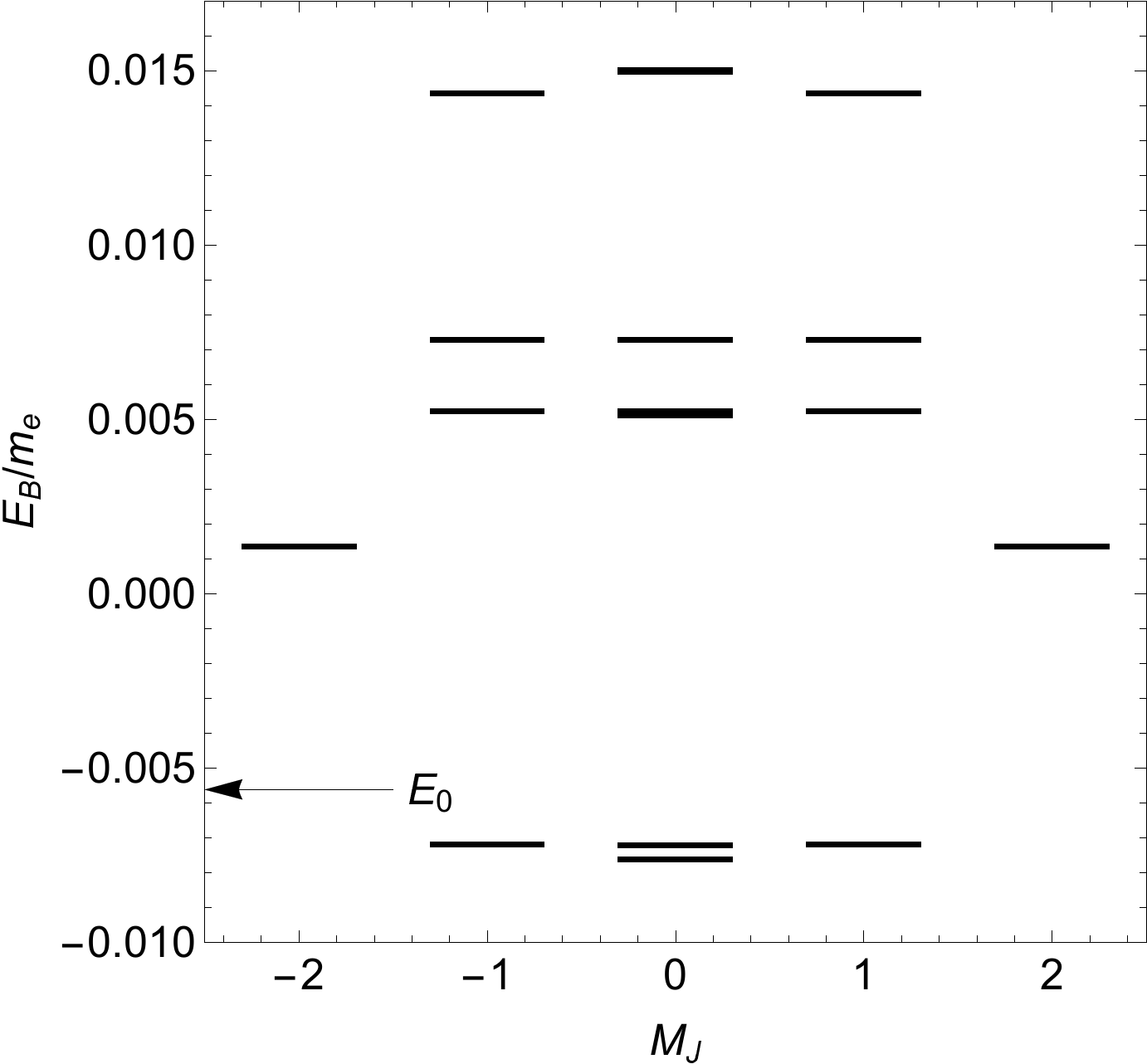}
\caption{\label{fig:massct} Left panel: representative value of the mass counterterm (in units of the physical electron mass $m_e$) in the positronium problem as a function of basis truncation parameters $N_{\rm{max}}=K-1$; right panel: the binding energy ($E_B$) spectrum of the positronium system at $N_{\rm{max}}=K-1=16$ and $\alpha=0.15$. $E_0$ is the ground state ($1^1S_0$) binding energy from nonrelativistic quantum mechanics with perturbative corrections. }
\end{figure}

In this calculation, we adopt the Fock-sector dependent renormalization~\cite{Karmanov:2008,Karmanov:2012,Zhao:2014hpa}, according to which only the fermion mass in the $|e^+ e^-\rangle$ sector needs to be renormalized, namely, the bare mass $m_{e0}$ is different from the physical mass $m_e$.  Different basis states take distinct values for the mass counterterm depending on their respective quanta available for self-energy fluctuation. The mass counterterms are determined from solving a series of single electron systems in the $|e\rangle+ |e \gamma\rangle$ Fock sectors~\cite{Zhao:2014xaa}.

\begin{figure}
\centering
\includegraphics[width=0.3\columnwidth
]{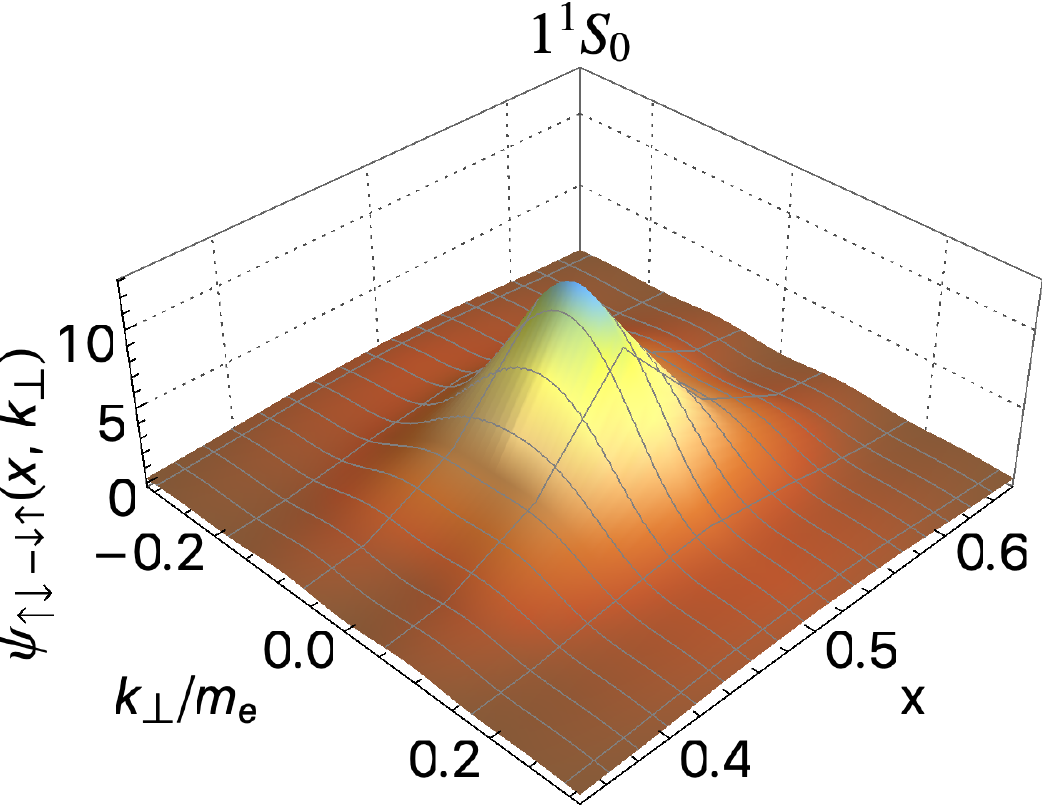}
\includegraphics[width=0.3\columnwidth
]{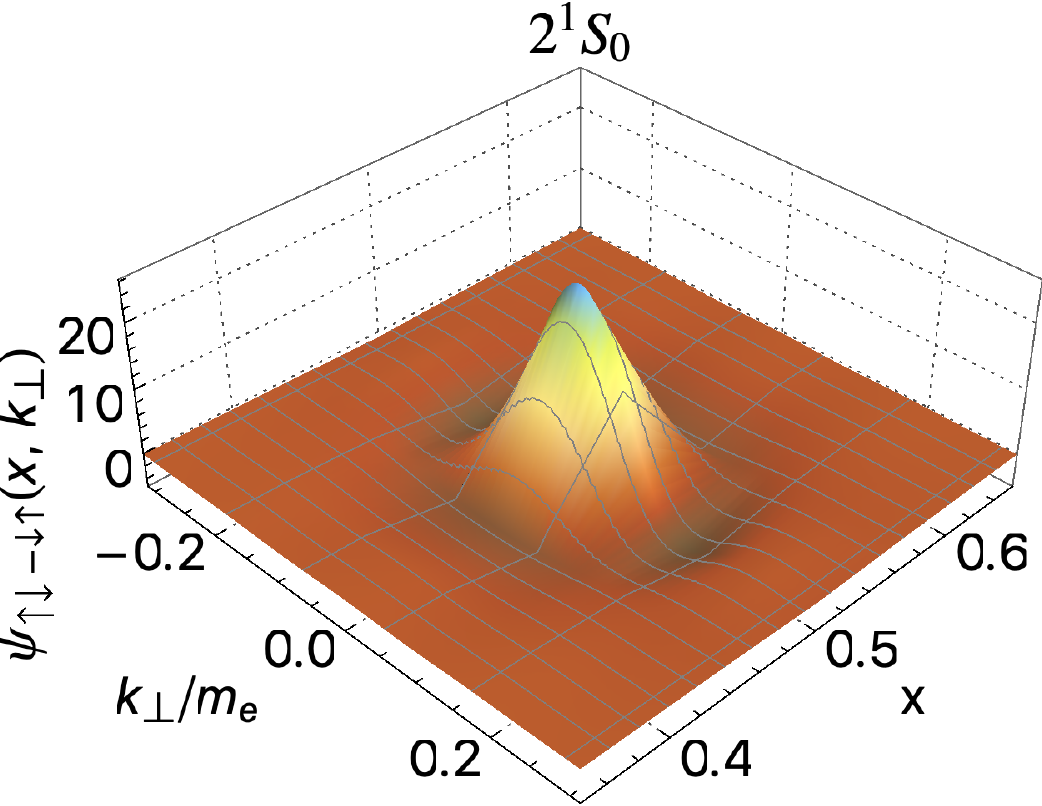}
\includegraphics[width=0.3\columnwidth
]{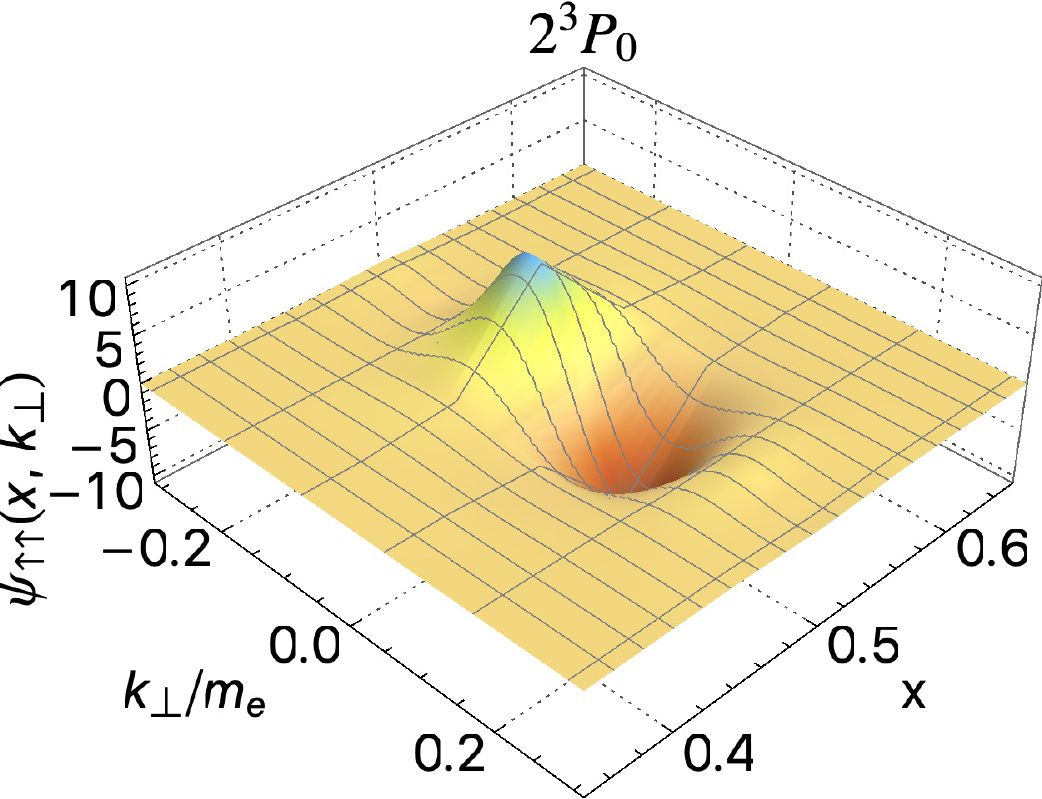}
\caption{\label{fig:wf} The (normalized) LFWFs for the dominant spin component in the $|e^+ e^-\rangle$ sector of the positronium system at $N_{\rm{max}}=K-1=16$ and $\alpha=0.15$. $x$ is the longitudinal momentum fraction and $k_\perp$ represents the relative transverse momentum between $e^+$ and $e^-$.}
\end{figure}

The value of the mass counterterm $\Delta m=m_{e0}-m_e$ for a representative basis state in the positronium problem as a function of the truncation parameters is shown in the left panel of Fig.~\ref{fig:massct} and in the right panel, we present the binding energy spectrum of the positronium system, $E_B \equiv M_{\rm{Ps}}-2\,m_e$, for different spin projections $M_J$. The arrow indicates the value of the ground state binding energy from nonrelativistic quantum mechanics~\cite{bs}, which is close to our result. We note that the mass counterterm is typically on a larger scale than that of the binding energy. We also note that the scale of the binding energy and structures of multiplets in the spectrum are in reasonable agreement with the previous calculation based on an effective one-photon-exchange interaction between the $e^+$ and $e^-$~\cite{Wiecki:2014}. The approximate degeneracy among different $M_J$ substates and the information from the mirror parity and charge parity~\cite{Li:2017mlw} allow us to identify the low-lying eigenstates. For the $M_J=0$ states, from bottom to up, the lowest six states are $1^1S_0$, $1^3S_1$, $2^1S_0$, $2^3S_1$, $2^3P_0$, and $2^3P_1$. Fig.~\ref{fig:wf} illustrates the light-front wave function (LFWF) in the $|e^+ e^-\rangle$ sector for three low-lying states with $M_J =0$. Their shape and nodal structures are qualitatively similar to those based on the one-photon-exchange effective interaction~\cite{Wiecki:2014}.

\section{Heavy meson}
A similar calculation can be carried over to the heavy meson system in QCD. Like the positronium system, we retain the lowest two Fock sectors, $|q\bar{q}\rangle$ and $|q\bar{q}g \rangle$, in our basis. Our Hamiltonian contains two parts. From the QCD Lagrangian, we obtain the first part of the Hamiltonian~\cite{Brodsky:1997de}, $P^-_{\rm{QCD}}$, which in our truncated Fock space takes a similar form to the QED Hamiltonian, $P^-_{\rm{QED}}$, in Eq.~(\ref{QEDHami}), with the fermion field $\Psi$ identified as the quark field, the gauge boson field $A_\mu$ identified as the gluon field $A^a_\mu T^a$ and the electric charge $e$ replaced by the color charge $g$. In order to achieve a more accurate reproduction of the meson mass spectrum, we allow the quark mass appearing in the quark-gluon vertex interaction, $gj^\mu A^a_\mu T^a$, to be an independent phenomenological parameter, $m'_q$, from $m_q$ in the kinetic energy term. We also apply a nonzero gluon mass $m_g$ to ensure the low-lying states are dominated by the $|q\bar{q}\rangle$ sector.

In addition we include a phenomenological confining potential~\cite{Li:2017mlw} in both the longitudinal and transverse directions in the $|q\bar{q}\rangle$ sector, which takes the following form,
\begin{equation}
P_{\rm C}^-P^+=\kappa^{4} \vec{\zeta}_{\perp} - \frac{\kappa^4}{(m_{q}+m_{\bar{q}})^2}\partial_{x}(x (1-x)\partial_{x}),
\end{equation}
where $x$ is the longitudinal momentum fraction of the quark~\cite{Wiecki:2014}. $\vec{\zeta}_{\perp}\equiv \sqrt{x(1-x)} \vec{r}_\perp$ is the holographic variable introduced by Brodsky and de T\'eramond~\cite{Brodsky:2014yha}, and $\partial_x f(x, \vec\zeta_\perp) = \partial f(x, \vec \zeta_\perp)/\partial x|_{\vec\zeta}$. $\kappa$ is the strength of the confinement and $m_{q}(m_{\bar{q}})$ is the mass of the quark (anti-quark). Thus, our total Hamiltonian is $P^-=P^-_{\rm QCD}+P^-_{\rm C}$.

\begin{figure}
\centering
\includegraphics[width=0.32\columnwidth
]{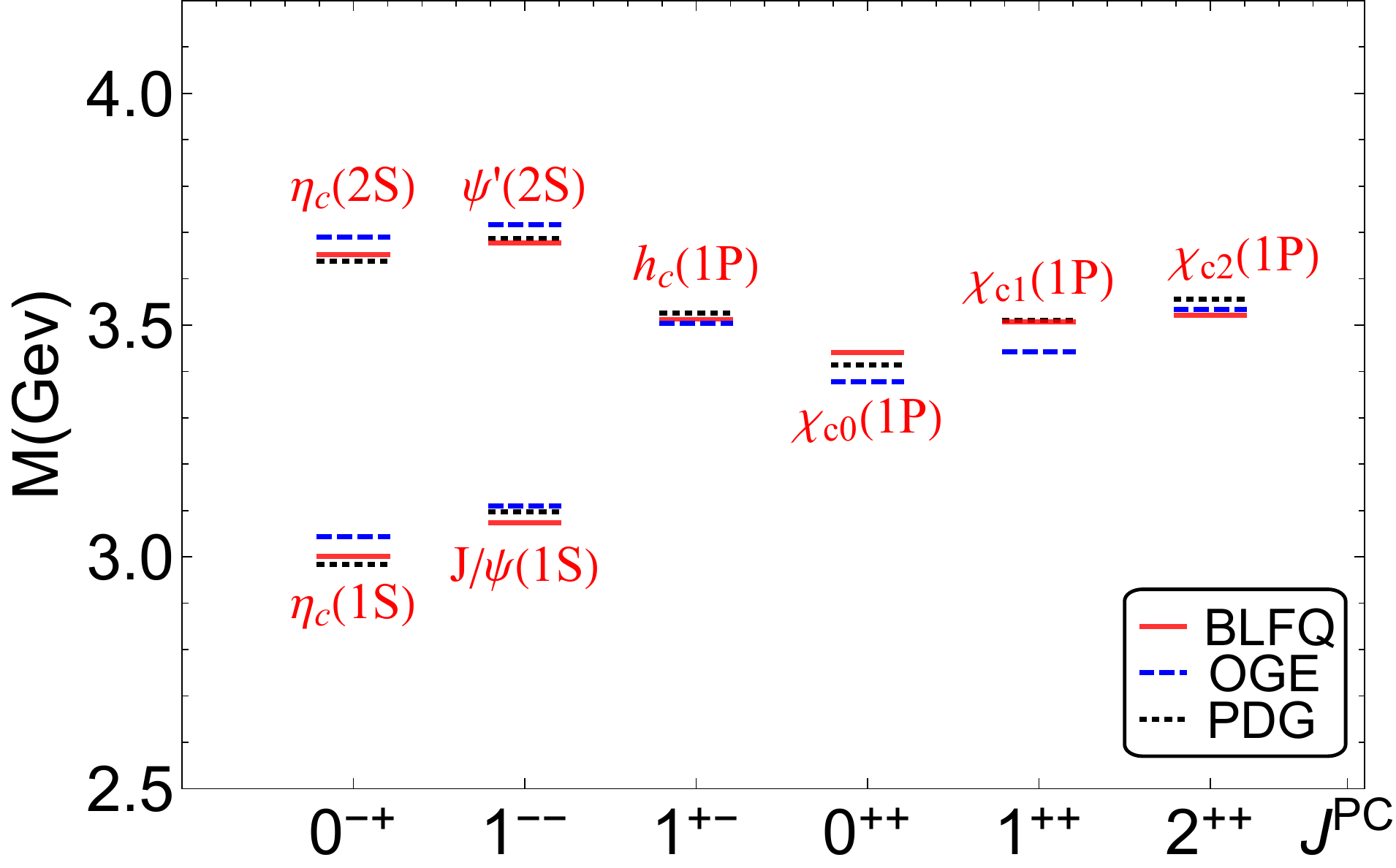}
\includegraphics[width=0.32\columnwidth
]{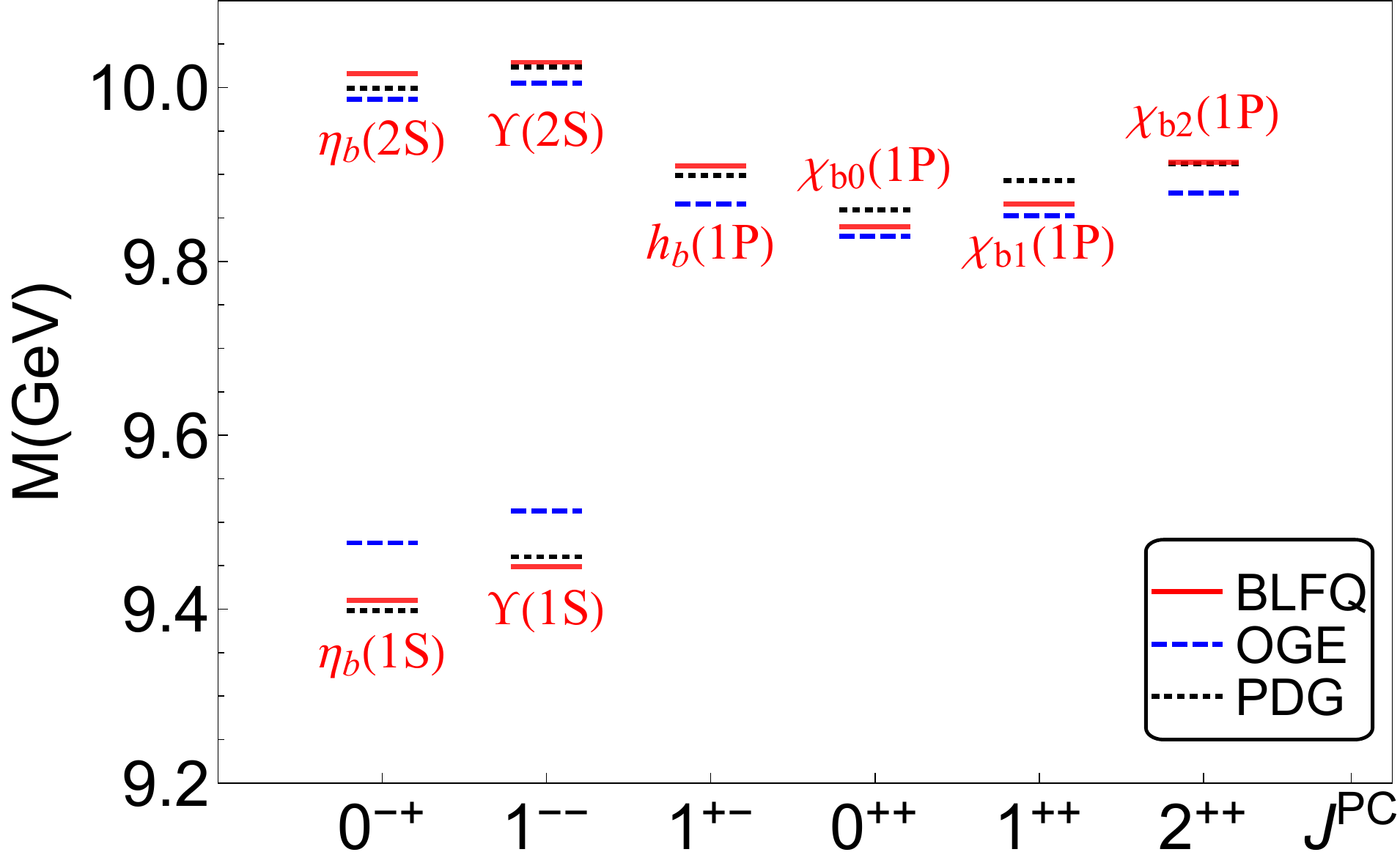}
\includegraphics[width=0.32\columnwidth
]{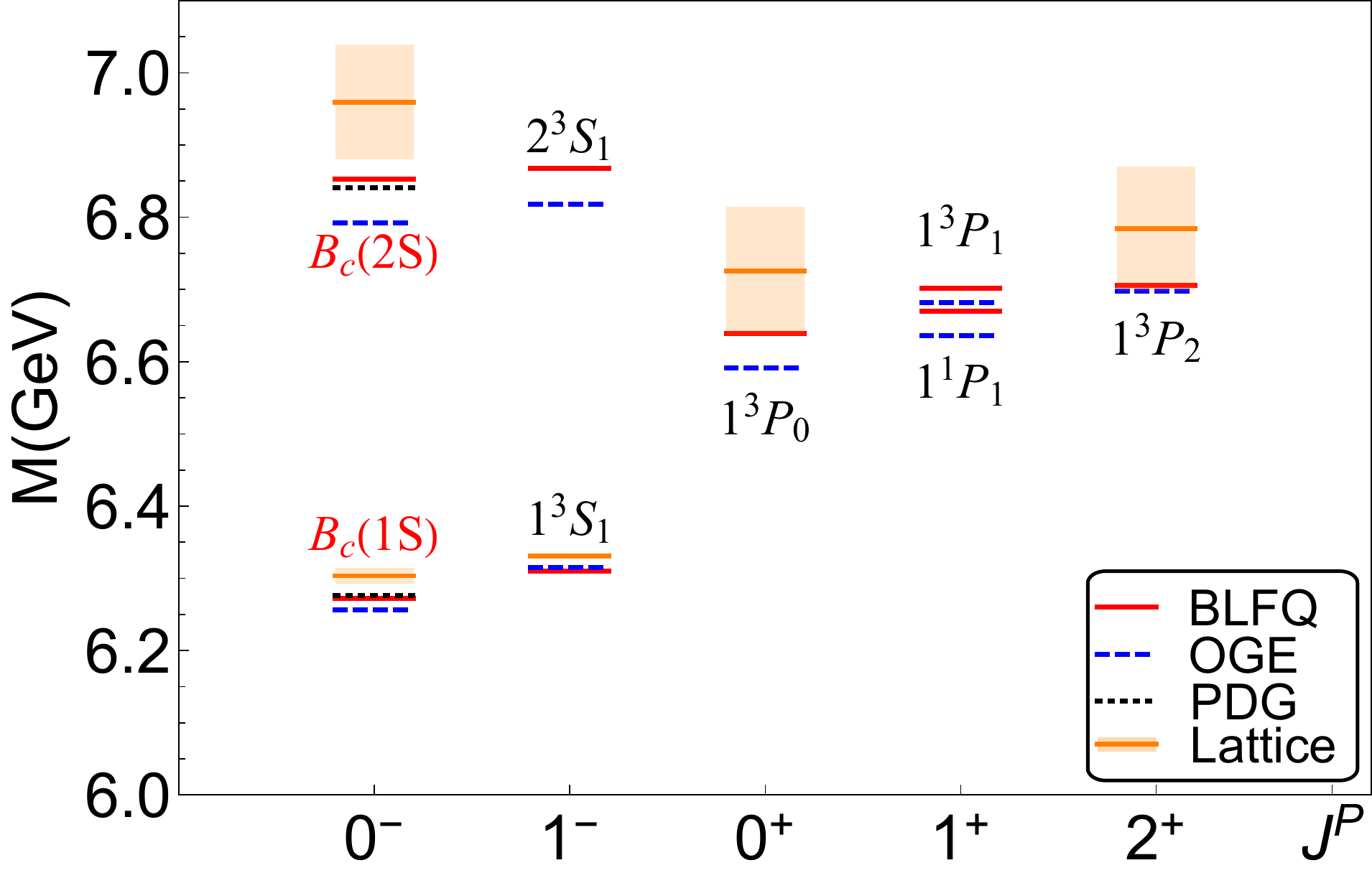}
\caption{\label{fig:en} Comparison of our BLFQ spectra at $N_{\rm{max}}=K-1=10$ for charmonium (left), bottomonium (middle), and $B_c$ meson (right) with the effective one-gluon-exchange (OGE) approach~\cite{Li:2017mlw,Tang:2018myz} and the experimental values (PDG)~\cite{Tanabashi:2018oca}. Lattice results are from Ref.~\cite{Allison:2004be,Gregory:2009hq,Davies:1996gi}. The horizontal and vertical axes are the $J^{\mathsf P\mathsf C}$ and invariant mass, respectively. }
\end{figure}
\begin{figure}
\centering
\includegraphics[width=0.3\columnwidth
]{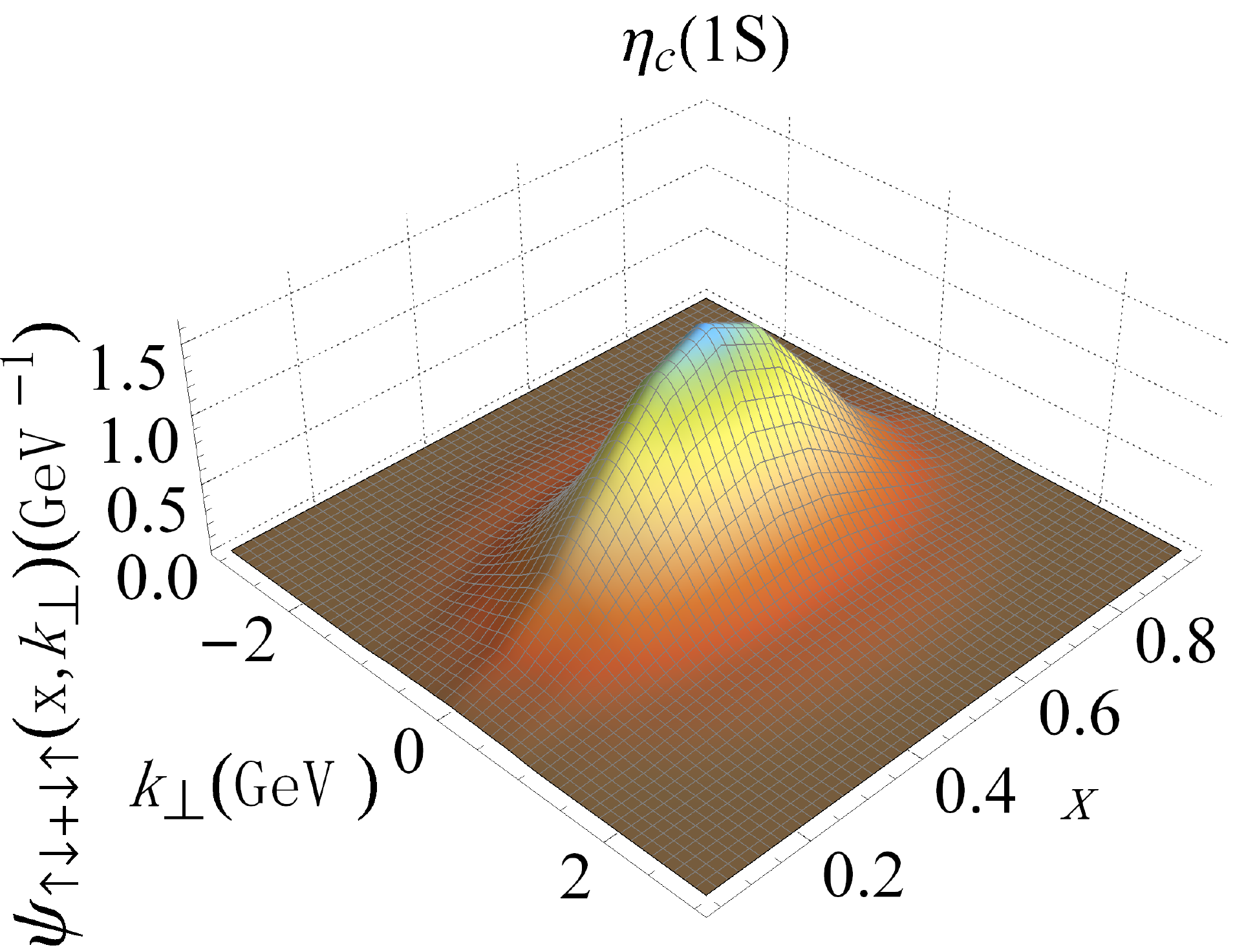}
\includegraphics[width=0.3\columnwidth
]{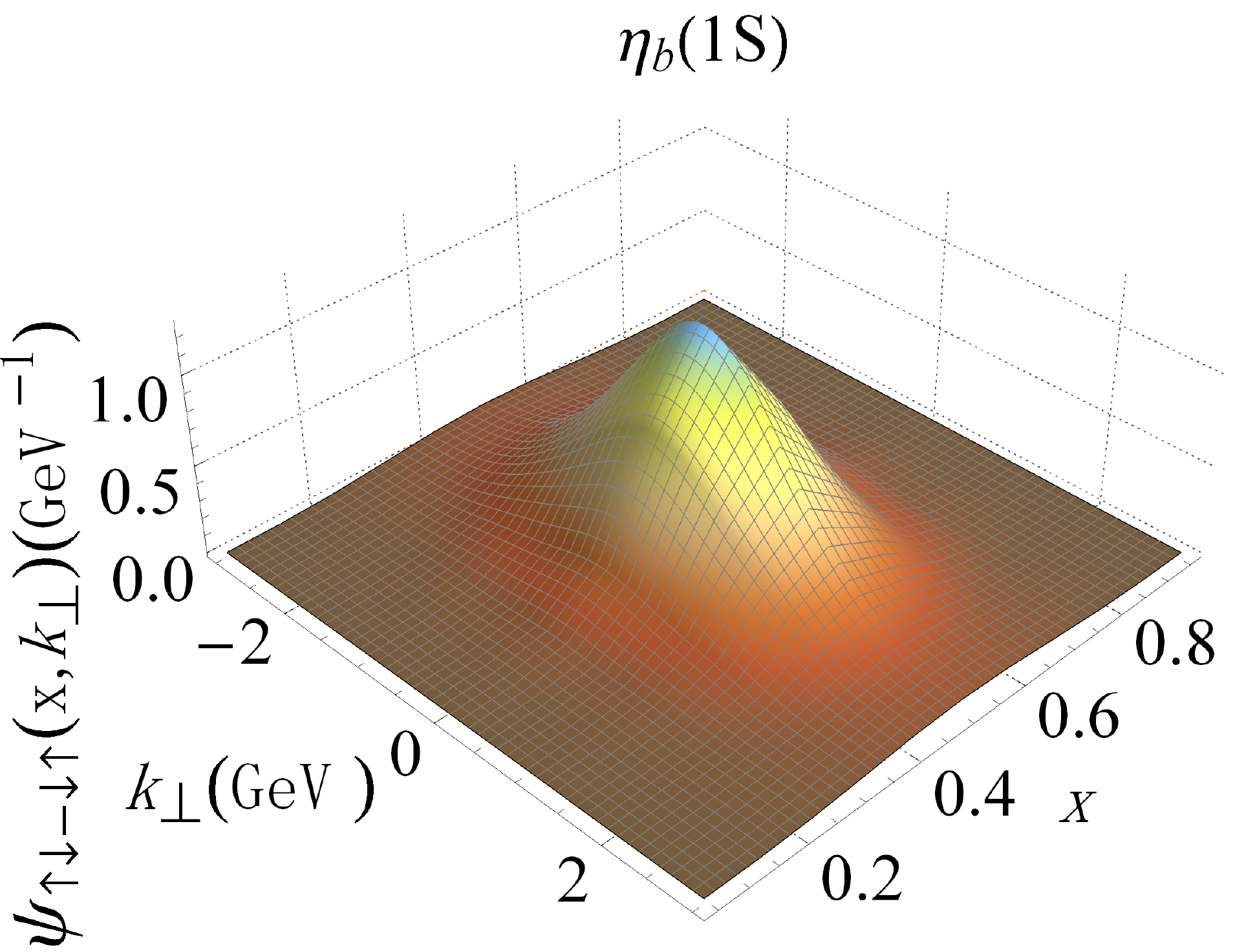}
\includegraphics[width=0.3\columnwidth
]{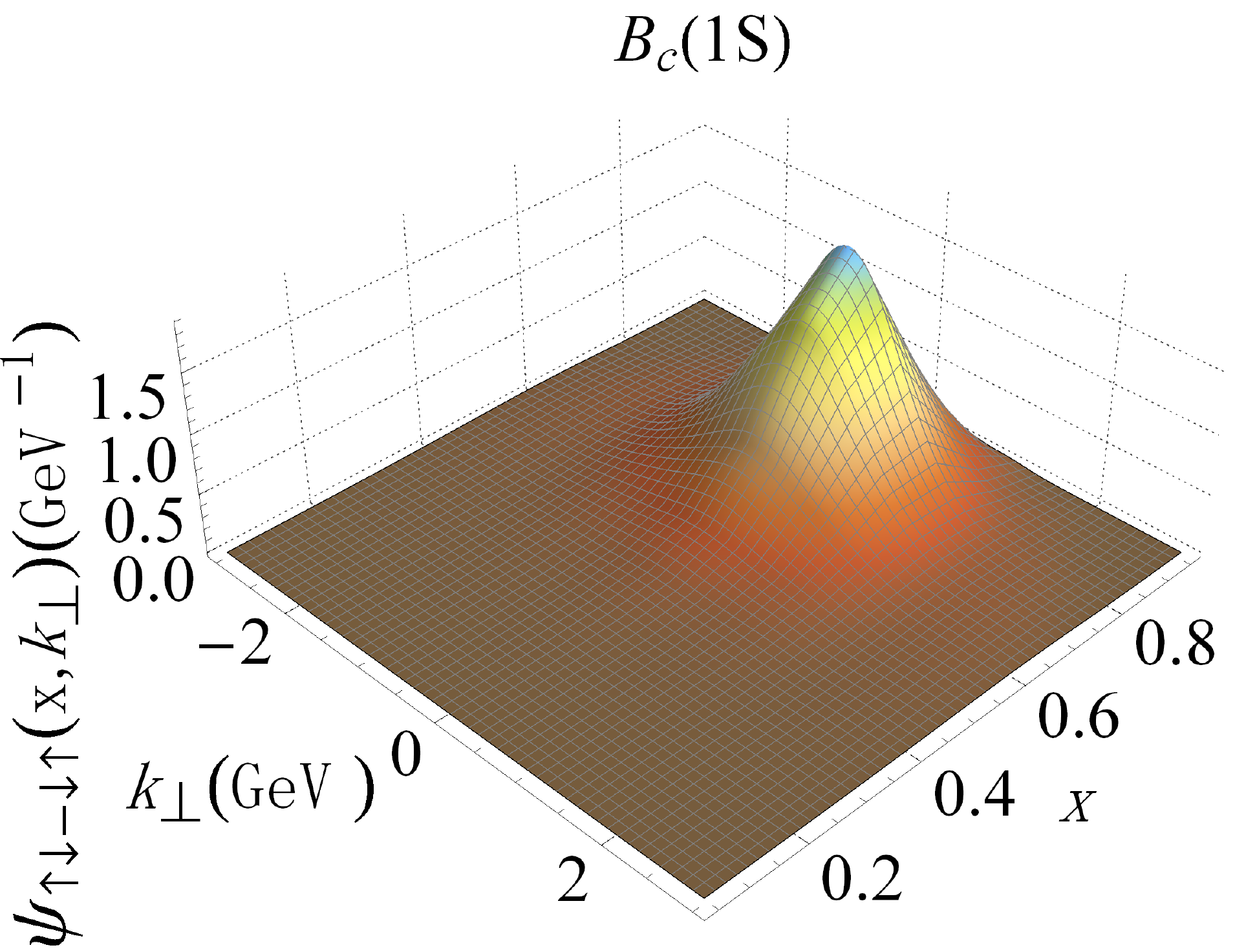}
\caption{\label{fig:wfq1} The (normalized) LFWFs for the dominant spin component in the $|q\bar{q}\rangle$ sector of $\eta_c(1S)$ (left), $\eta_{b}(1S)$ (middle) and $B_c(1S)$ (right) at $N_{\rm{max}}=K-1=10$. $x$ is the longitudinal momentum fraction of the quark and $k_\perp$ represents the relative transverse momentum between the quark and the anti-quark.}
\end{figure}

With the charm quark mass $m_c=1.565$\,GeV (kinetic), $m'_c=6.259$\,GeV (interaction), the gluon mass $m_g=0.5$\,GeV, the confining strength $\kappa=1.117$\,GeV and the strong coupling constant $g=1.51$ for the charmonium, and similarly the bottom quark mass $m_b=4.767$\,GeV (kinetic), $m'_b=12.191$\,GeV (interaction), $m_g=0.5$\,GeV, $\kappa=1.894$\,GeV and $g=1.75$ for the bottomonium, our resulting mass spectra for the low-lying $c\bar{c}$ and $b\bar{b}$ states agree with the experimental values reasonably well, as shown in the left and middle panel of Fig.~\ref{fig:en}, respectively. With the same parameter set used in the $c\bar{c}$ and $b\bar{b}$ systems we obtain the mass spectrum for the $B_c$ system, as in the right panel of Fig.~\ref{fig:en}. 
In Fig.~\ref{fig:wfq1} we present the ground state LFWFs of the $c\bar{c}$, $b\bar{b}$ and $B_c$ system in the $|q\bar{q}\rangle$ sector.

\section{Conclusion}
Through the applications to various bound state systems in QED and QCD we demonstrate that BLFQ is a versatile and powerful nonperturbative approach to quantum field theory for strongly interacting systems. We anticipate that in the future BLFQ will be a useful tool for understanding the hadron mass spectrum and structure beyond the valence sector.

{\it Acknowledgment:}
XZ is supported by Key Research Program of Frontier Sciences, CAS, Grant No ZDBS-LY-7020. JPV is supported by the Department of Energy under Grants No. DE-FG02-87ER40371, and No. DE-SC0018223 (SciDAC4/NUCLEI). A portion of the computational resources were provided by the National Energy Research Scientific Computing Center (NERSC), which is supported by the Office of Science of the U.S. Department of Energy under Contract No.DE-AC02-05CH11231.


\begin{thebibliography}{99}

\bibitem{Vary:2009gt}
  J.~P.~Vary {\it et al.},
  Phys.\ Rev.\ C {\bf 81}, 035205 (2010).


\bibitem{Fu:2020b}
  K.~Fu, H.~Zhao, X.~Zhao, J.~P.~Vary,
  to be published in the proceedings of Hadron 2019 conference (2020).

\bibitem{Karmanov:2008}
  V.~A.~Karmanov, J.-F.~Mathiot and A.~V.~Smirnov,
  Phys.\ Rev.\ D {\bf 77}, 085028 (2008).

\bibitem{Karmanov:2012}
  V.~A.~Karmanov, J.~F.~Mathiot and A.~V.~Smirnov,
  Phys.\ Rev.\ D {\bf 86}, 085006 (2012).

\bibitem{Zhao:2014hpa}
  X.~Zhao,
  Few Body Syst.\  {\bf 56}, no. 6-9, 257 (2015).

\bibitem{Zhao:2014xaa}
  X.~Zhao, H.~Honkanen, P.~Maris, J.~P.~Vary and S.~J.~Brodsky,
  Phys.\ Lett.\ B {\bf 737}, 65 (2014).

\bibitem{bs} H. A. Bethe and E. E. Salpeter, {\it Quantum Mechanics of One- and Two-Electron Atoms},
Springer, Heidelberg, 1957.

\bibitem{Wiecki:2014}
  P.~Wiecki, Y.~Li, X.~Zhao, P.~Maris and J.~P.~Vary,
  Phys.\ Rev.\ D {\bf 91}, no. 10, 105009 (2015).


  \bibitem{Brodsky:1997de}
  S.~J.~Brodsky, H.~C.~Pauli and S.~S.~Pinsky,
  Phys.\ Rept.\  {\bf 301}, 299 (1998).

\bibitem{Li:2017mlw}
  Y.~Li, P.~Maris and J.~P.~Vary,
  Phys.\ Rev.\ D {\bf 96}, no. 1, 016022 (2017).

\bibitem{Brodsky:2014yha}
  S.~J.~Brodsky, G.~F.~de Teramond, H.~G.~Dosch and J.~Erlich,
  Phys.\ Rept.\  {\bf 584}, 1 (2015).

\bibitem{Tang:2018myz}
S.~Tang, Y.~Li, P.~Maris and J.~P.~Vary,
Phys.\ Rev.\ D {\bf 98}, no. 11, 114038 (2018).


\bibitem{Tanabashi:2018oca}
M.~Tanabashi {\it et al.} [Particle Data Group],
Phys.\ Rev.\ D {\bf 98}, no. 3, 030001 (2018).


\bibitem{Allison:2004be}
  I.~F.~Allison {\it et al.} [HPQCD and Fermilab Lattice and UKQCD Collaborations],
  Phys.\ Rev.\ Lett.\  {\bf 94}, 172001 (2005).

\bibitem{Gregory:2009hq}
  E.~B.~Gregory {\it et al.},
  Phys.\ Rev.\ Lett.\  {\bf 104}, 022001 (2010).

\bibitem{Davies:1996gi}
  C.~T.~H.~Davies, K.~Hornbostel, G.~P.~Lepage, A.~J.~Lidsey, J.~Shigemitsu and J.~H.~Sloan,
  Phys.\ Lett.\ B {\bf 382}, 131 (1996).













\end{thebibliography}
\end{document}